\documentclass[fleqn,usenatbib]{mnras}

\usepackage{newtxtext,newtxmath}

\usepackage[T1]{fontenc}
\usepackage{ae,aecompl}

\usepackage{graphicx}	
\usepackage{amsmath}	
\usepackage{amssymb}	

\usepackage{longtable}

\title[Redshifts of Infrared-Faint Radio Sources]{The Redshift Distribution of Infrared-Faint Radio Sources}

\author[Brendan J. Orenstein et al.]{
Brendan J. Orenstein,$^{1,2}$
Jordan D. Collier,$^{1,3,4}$
Ray P. Norris$^{1,3}$\thanks{E-mail:raypnorris@gmail.com}
\\
$^{1}$CSIRO Astronomy and Space Science (CASS), Marsfield, NSW 2122, Australia\\
$^{2}$Research School of Astronomy \& Astrophysics, The Australian National University, Cotter Road, Weston Creek, ACT 2611, Australia\\
$^{3}$Western Sydney University, Locked Bag 1797, Penrith, NSW 2751, Australia\\
$^{4}$The Inter-University Institute for Data Intensive Astronomy (IDIA), Department of Astronomy, University of Cape Town, Rondebosch, 7701, South Africa\\
}

\date{Accepted XXX. Received YYY; in original form ZZZ}

\pubyear{2018}

\begin{document}
\label{firstpage}
\pagerange{\pageref{firstpage}--\pageref{lastpage}}
\maketitle

\begin{abstract}
Infrared-Faint Radio Sources (IFRSs) are an important class of high-redshift active galaxy, and potentially important as a means of discovering more high-redshift radio sources, but only 25 IFRSs had redshifts prior to this paper. Here we increase the number of IFRSs with known spectroscopic redshifts by a factor of about 5 to 131, with redshifts up to $z=4.387$, and a median redshift of z = 2.68. The IFRS redshift distribution overlaps with the high-$z$ radio galaxy (HzRG) redshift distribution but is significantly narrower, suggesting that the IFRSs are a subset of the larger class of HzRGs.
We also confirm and measure the proposed correlation between redshift and 3.6$\mu$m flux density, making it possible to use this correlation to find even higher redshift radio sources.
Many more high-redshift sources are probably present in existing radio survey catalogues.
\end{abstract}




\section{Introduction}

Infrared-faint radio sources (IFRSs) are galaxies that are strong at radio wavelengths and weak in the infrared. Most are at high redshift, and selecting them represents a valuable technique for finding high-redshift radio sources. Although this class of objects is well-studied \citep{norris06, norris07, middelberg08, middelberg10, huynh10, zinn11, norris11, collier14, garn08, herzog14, herzog15, herzog15a, herzog16, maini16, singh14, singh17}, their faintness at optical/IR wavelengths means that only 25 of them have measured spectroscopic redshifts, all but one of which are at redshift z $>$ 2.

~\cite{norris06} first identified this class of source by cross-matching observations from the Australia Telescope Large Area survey (ATLAS) survey and the Spitzer Wide-Area Extragalactic (SWIRE;~\citealt{lonsdale03}) survey. Out of the 2002 radio sources found by ATLAS in the Chandra Deep Field South (CDFS;~\citealt{rosati02}) and European Large Area ISO Survey - South 1 (ELAIS-S1;~\citealt{oliver00}) fields, only 53 were IFRSs, making them rare.

~\cite{norris11} suggested that IFRSs were most likely radio-loud active galactic nuclei (AGN) at high redshifts.
They also considered the less likely possibility of IFRSs being radio-loud AGN at a redshift of $1 < z < 3$, with dust extinction reducing the luminosity.
~\cite{middelberg10} compared the ratio of the 1.4 GHz and 3.6$\,\mu$m flux densities for an IFRS sample, a High-$z$ Radio Galaxy (HzRG) sample and a general radio source population sample (constructed from ATLAS catalogues), demonstrating 
that this ratio was common to the classes of IFRSs and HzRGs but not to the class of general radio sources.

~\cite{norris07} detected an IFRS using Very Long Baseline Interferometry (VLBI), showing that the source was probably an AGN, since a VLBI detection implies brightness temperatures greater than $10^6$ K, which can only be generated by an AGN.
~\cite{middelberg08} imaged an IFRS using VLBI, showing that the size, spectrum, radio and infrared luminosity were consistent with the properties of a high redshift compact steep-spectrum (CSS) source and inconsistent with the properties of a low redshift galaxy of low luminosity or a normal radio galaxy.
\cite{herzog15} used VLBI observations to show that the majority of IFRSs probably contain AGN.

While \cite{norris06} defined criteria for IFRSs that depended on survey sensitivity, \cite{zinn11} proposed generalised criteria that are survey-independent, so that they can be applied  to all astronomical surveys. The \cite{zinn11} criteria are:
\begin{itemize}
\item A flux density ratio of $S_{\rm 20\,cm}/S_{\rm 3.6\,\mu m} > 500$ 
\item A 3.6$\,\mu$m flux density of $< 30\,\mu$Jy
\end{itemize}
The first criterion selects sources with extreme infrared to radio flux density ratios, similar to HzRGs. The second criterion removes radio-loud AGN with low redshifts.

Using these criteria, \cite{collier16} generated a sample of 1317 IFRSs, which included 93 compact steep-spectrum (CSS) sources and 31 GHz peaked-spectrum (GPS) sources. CSS are compact, powerful radio sources with a spectral peak at $\sim$100 MHz, while GPS have a spectral peak at $\sim$1 GHz ~\citep{orienti15}. Using a sample of 14 IFRSs from the ELAIS-S1 and 14 from the CDFS,~\cite{herzog16} determined that $\geq15^{+8}_{-4}$ per cent of their sample were CSS sources and $3^{+6}_{-1}$ per cent were GPS sources. 
\cite{singh17} determined the radio morphologies of a sample of 11 IFRSs from the Subaru X-ray Deep Field (SXDF) and 8 from the Very Large Array - VIMOS VLT Deep Survey (VLA-VVDS) field. Of the total sample of 19 IFRSs, 14 were unresolved point sources and five featured extended double-lobed morphologies, classifying them as radio galaxies.

Finding high-redshift radio sources is important both as probes of the intergalactic medium, and as a means of studying the sub-kpc morphology of active galactic nuclei in the early Universe, including the possibility of detecting binary supermassive black holes. Previous attempts to find high-redshift radio sources have mainly used the apparent correlation between redshift and spectral index \citep[e.g.][]{miley08}. Since virtually all IFRS are at z $ > $ 2, they are potentially an even more efficient technique for finding  high-redshift sources.

Determining the redshift of IFRSs is key to understanding their nature. Because of their faintness, very few redshifts have been measured. ~\cite{herzog14} determined spectroscopic redshifts for three sources of $z$ = 1.84, 2.13 and 2.76. 
Of the 1317 IFRSs examined by~\cite{collier14} from the Wide-field Infrared Survey Explorer ~\citep[WISE;][]{wright10} All-Sky data release \citep{cutri12}, only 19 had spectroscopic redshifts listed in SDSS DR9. 
Remarkably, 18 of these had redshifts of z $\geq$ 2, suggesting that the IFRSs with spectroscopic redshift measurements are unlikely to be nearby AGN, but are more likely to be high-redshift AGN. As the majority of WISE sources have redshifts of $z<1$, these sources were very unlikely to be misidentifications. ~\cite{herzog15} also determined the photometric redshifts of 11 IFRSs observed with the Very Long Baseline Array. Three of these had spectroscopic redshift measurements listed in the Sloan Digital Sky Survey (SDSS) Data Release 10 (DR10) of $z$ = 2.11, 2.55 and 2.62, which had already been listed by \cite{collier14}. Of the 19 IFRSs identified by~\cite{singh17}, only three had spectroscopic redshifts, which were $z$ = 2.43, 2.47 and 3.57. Obtaining spectroscopic redshifts for a larger sample of IFRSs is crucial to determine their true nature.

In this paper we do not distinguish between high-redshift radio galaxies and radio-loud quasars, which differ intrinsically only by their orientation \citep{urry95}, as we are primarily concerned with the flux density at 20 cm and 3.6 $\mu \mathrm{m}$. At mJy sensitivities, the flux of most radio galaxies is dominated by hotspot and lobe emission, and so (with the exception of the relatively rare flat-spectrum quasars and blazars) is independent of orientation, and so radio galaxies and quasars with similar host properties have similar flux densities \citep[e.g.][]{urry95}. 

The orientation dependence at 3.6 $\mu \mathrm{m}$ is less well-defined, since this corresponds to a rest wavelength of 1.8--0.7 $\mu \mathrm{m}$ for the redshift range z $\sim$ 1--4 of our sample. At these wavelengths, the emission will contain contributions from the accretion disc, dusty torus, and the host galaxy, with the importance of the accretion disc, and extinction by the torus, increasing at shorter wavelengths \citep{hernan16}, leading to some orientation dependence.   However, the optical spectra, at even shorter wavelengths,  are even more strongly affected by the orientation, making quasars more easily observable in spectroscopy than radio galaxies. Therefore,  IFRSs with measured redshifts are much more likely to be quasars than IFRSs without measured redshifts. As a result, the majority of IFRSs discussed in this paper, and in papers such as \cite{herzog14}, are quasars. However their radio and infrared properties are similar to those of radio galaxies, which probably constitute the majority of the IFRSs.

The goal of this paper is to obtain a larger sample of IFRSs with spectroscopic redshifts. Section 2 of this paper introduces our dataset. Section 3 presents our redshift distribution. Section 4 discusses the redshift distribution and Section 5 presents our conclusions.

\section{Data}

Our data were selected following a procedure similar to \cite{collier14}, using deeper photometry from WISE and more extensive spectroscopy from SDSS.

Our IFRS sample was taken from Version 2.0 of the Unified Radio Catalog ~\citep[URC;][]{kimball08,kimball14}. This catalog contains approximately three million radio sources, and combines the 20 cm flux density measurements from the Faint Images of the Radio Sky at Twenty Centimeters (FIRST;~\citealt{becker95}) survey and the NRAO-VLA Sky Survey (NVSS;~\citealt{condon98}). 

We obtained infrared data from the AllWISE data release \citep{cutri14}, which improves upon the sensitivity, photometric and astrometric accuracy of the WISE All-Sky data release. We converted 3.4$\,\mu$m measurements from magnitudes to Jansky (Jy) with a flat color correction factor, using the conversion determined for IFRSs by \cite{collier14}:
\begin{equation}
    S_{3.4\,{\mu}\mathrm{m}}=306.682\times10^{(-M_{3.4\,\mu \mathrm{m}} /2.5)}Jy
    \label{eq:conversion}
\end{equation}

We used spectroscopic redshifts from SDSS DR12 \citep{alam15}. DR12 contains measurements from 2008 to 2014, taken by the third generation of SDSS.

\subsection{Selection Criteria}

\begin{figure*}
    \centering
    \includegraphics[width=\textwidth]{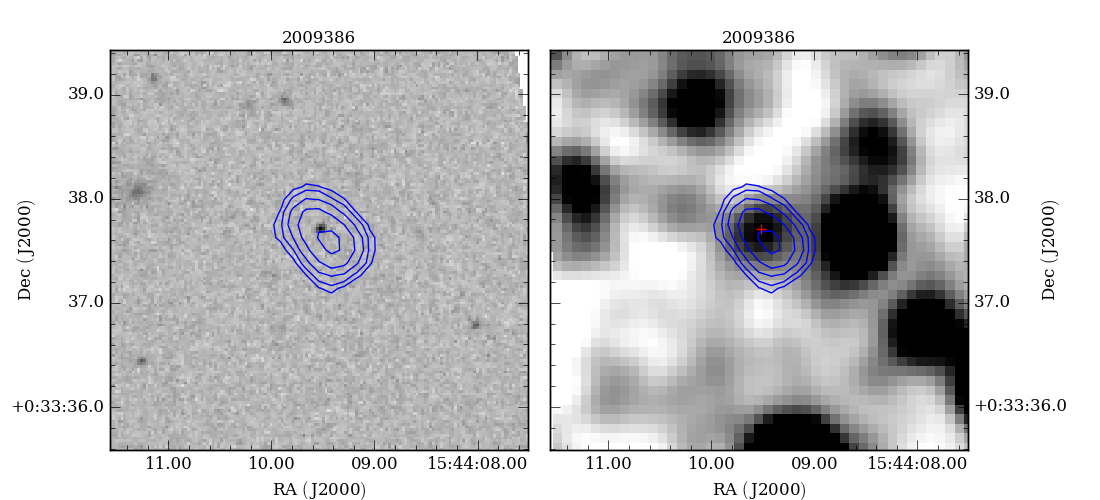}
    \caption{An example postage stamp of an IFRS, generated for visual inspection. FIRST contours are shown in blue, representing 4-1024 times the local r.m.s., increasing by multiples of 2, overlaid on the the SDSS $i$ band image (left), and the AllWISE 3.4$\,\mu$m image (right). The red cross in the right image shows the position of the SDSS fibre used to measure the redshift. Both panels are labelled according to their unique ID from the URC.
}
    \label{fig:postageStamp}
\end{figure*}

In Table~\ref{table:selectionCriteria}, we present the  selection criteria followed to select our sample, and the number of sources remaining after applying each criterion, which we now explain in more detail:

\begin{enumerate}
  \setcounter{enumi}{-1}
  \item We began with the 2,866,856 sources from the URC v2.0. 
  \item We applied a NVSS flux density limit of $S_{\rm 20\,cm}$ > 7.5 mJy. This was applied to reduce the size of the following WISE query, since all IFRSs with $S_{\rm 20\,cm}/S_{\rm 3.4\,\mu m} > 500$ will have $S_{\rm 20\,cm}$ > 7.5 mJy, following a 5$\sigma$ detection and an r.m.s. of $\sigma > 3\,\mu$Jy at 3.4 $\,\mu$m.
  \item We removed all sources that did not have at least one FIRST counterpart, which we used for the radio source positions, as FIRST has a higher angular resolution than NVSS.
  \item We cross-matched the FIRST positions to AllWISE. We used a search radius of 5 arcsec to ensure minimal false matches at redshifts of $z > 0.5$.
  \item We applied the second of~\cite{zinn11}'s generalised IFRS criteria, $S_{\rm 20\,cm}/S_{\rm 3.4\,\mu m} > 500$.
  \item We applied a 3.4$\,\mu$m signal-to-noise ratio (SNR) limit of $\geq5$.
  \item We cross-matched our sources to SDSS DR12 using the AllWISE positions and a search radius of 1 arcsec.
  \item We selected all sources with spectroscopic redshifts.
  \item We removed all duplicate sources that shared the same SDSS optical spectroscopic object identification number.
  \item We removed all sources with a spectral classification of a STAR, ensuring all sources were extragalactic.
  \item We removed all sources with redshift warning flags.
  \item We removed all sources with negative redshift errors, as for these sources, the errors could not be determined.
  \item We selected all sources that had Q (observation quality) values of 3 for good, as opposed to 1 for bad or 2 for acceptable. 
  \item We applied the first of ~\cite{zinn11}'s generalised IFRS criteria, $S_{\rm 3.4\,\mu m} < 30\,\mu$Jy, resulting in our IFRS sample.
  \item We visually inspected an image of each IFRS, using the AllWISE 3.4$\,\mu$m and SDSS $i$ band images and FIRST contours, such as the one shown in Figure~\ref{fig:postageStamp}. No sources were identified as misidentifications. 
\end{enumerate}
After we applied the first 12 selection criteria, our sample consisted of 2521 radio sources with spectroscopic redshifts from SDSS. We call this our ``initial large sample''. After applying all 14 criteria, our sample consisted of  108 IFRSs, and we call this our ``IFRS sample''. Two of these sources already had  spectroscopic redshifts listed by \cite{collier14}.

\begin{table}
\caption{The selection criteria and number of sources after applying each criterion to our IFRS sample.}
\label{table:selectionCriteria}
\begin{center}
 \begin{tabular}{||c c c||} 
 \hline
 No. & Selection Criterion & Sources \\ [0.5ex] 
 \hline\hline
 0 & Total Unified Radio Catalog & 2,866,856 \\ 
 1  & NVSS flux density $S_{\rm 20\,cm} > 7.5$ mJy & 1,139,132 \\
 2 & At least one FIRST counterpart & 621,316  \\
 3 & AllWISE match within 5$\arcsec$ of FIRST & 303,043 \\
 4 & $S_{\rm 20\,cm}/S_{\rm 3.4\,\mu m} > 500$ & 64826 \\
 5 & $S_{\rm 3.4\,\mu m}$ SNR $>= 5$ & 63998 \\
 6 & SDSS match within 1$\arcsec$ of AllWISE & 46490 \\
 7 & SDSS source with Spectroscopic Redshift & 5761 \\
 8 & Remove SDSS duplicates & 2798 \\
 9 & Not a star & 2747 \\
 10 & No zWarning flag & 2566 \\
 11 & Positive z Error & 2551 \\
 12 & Good quality observation & 2521 \\
 13 & $S_{\rm 3.4\,\mu m} < 30\,\mu$Jy & 108 \\
 14 & Visual Inspection of images and spectra & 108 \\
 \hline
\end{tabular}
\end{center}
\end{table}

\subsection{Positional Offsets}

\begin{figure}
	\includegraphics[width=\columnwidth]{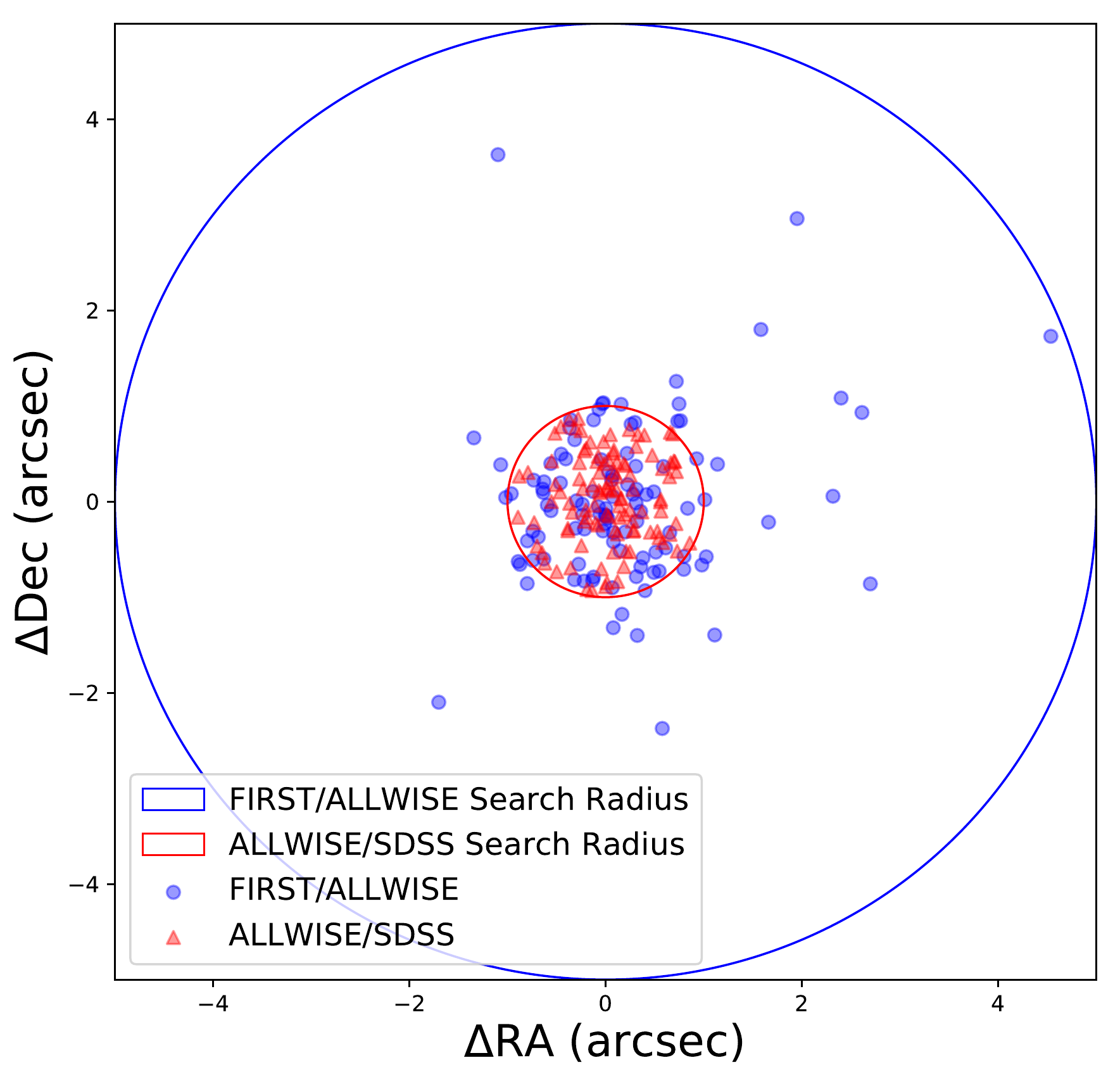}
    \caption{The sky separations for our 108 IFRSs with spectroscopic redshifts, between FIRST and AllWISE (blue circles) and AllWISE and SDSS (red triangles). The FIRST--AllWISE search radius of 1$\arcsec$ is shown in red, and the AllWISE--SDSS search radius of 5$\arcsec$ is shown in blue.}
    \label{fig:PositionalOffsets}
\end{figure}

We show positional offsets of the  sources in our IFRS sample between FIRST and ALLWISE and between ALLWISE and SDSS
in Figure~\ref{fig:PositionalOffsets}, and  display their median and mean values in Table~\ref{table:positionalOffsets}. This shows there are no significant systematic offsets.

\begin{table}
\caption{The positional offsets in arcsec between our datasets}
\label{table:positionalOffsets}
\begin{center}
 \begin{tabular}{||c c c c||} 
 \hline
  & Radial Separation & RA & Dec \\ [0.5ex] 
 \hline\hline
 FIRST--AllWISE Median & 0.87 & 0.58 & 0.47 \\
 FIRST--AllWISE Mean & 1.05 & 0.74 & 0.60 \\
 AllWISE--SDSS Median & 0.59 & 0.30 & 0.34 \\
 AllWISE--SDSS Mean & 0.61 & 0.39 & 0.38 \\
 \hline
\end{tabular}
\end{center}
\end{table}

\subsection{Misidentification Rates}
To estimate the misidentification rate, particularly due to confusion within the AllWISE catalogue, we downloaded a subset of the URC, AllWISE and SDSS catalogues from two regions of sky, given the non-uniformity of these surveys. Region one was constrained to one degree either side of a position of (5,5) degrees (i.e. 00:16:00 $\leq$ RA $\leq$ 00:24:00, and $4\deg \leq$ Dec $\leq 6\deg$), while region two was constrained to one degree either side of a position of (10,10) degrees (i.e. 00:36:00 $\leq$ RA $\leq$ 00:44:00, and $9\deg \leq$ Dec $\leq 11\deg$). We shifted the positions of all of the FIRST sources by a single random amount between either $-$60 to $-$20 arcsec or 20 to 60 arcsec. These shift sizes were chosen to be greater than the beam sizes for the FIRST, AllWISE and SDSS surveys. We then cross-matched FIRST to AllWISE with a search radius of 5$\arcsec$. 
We applied the selection criteria shown in Table 1 to the shifted sources. For computational reasons, the criteria were applied in a slightly different order than had been used on the original sample.
Lastly, we cross-matched AllWISE to the SDSS spectroscopic redshift catalogue of that region, with a search radius of 1$\arcsec$. This process was repeated 1000 times, each time shifting all FIRST positions by a different random amount.

We determined the misidentification rate by dividing the mean number of shifted cross-matches by the number of unshifted cross-matches. The result is shown in Table~\ref{table:misidentificationRates}, which shows that, after all criteria have been applied, essentially none of the selected sources, in either the large sample or the IFRS sample, are erroneous.

\begin{table}
\caption{The misidentification rates for survey cross-matches}
\begin{center}
 \begin{tabular}{||llll||} 
 \hline
 Selection Criteria & Rate (\%) & Rate (\%)\\
    & region 1 & region 2 \\ [0.5ex] 
 \hline\hline
 AllWISE match within 5$\arcsec$ of FIRST & 18.2 & 18.1\\
 NVSS flux density $S_{\rm 20\,cm} > 7.5$ mJy & 5.5 & 5.7 \\
$S_{\rm 20\,cm}/S_{\rm 3.4\,\mu m} > 500$ & 1.6 & 2.2 \\
$S_{\rm 3.4\,\mu m}$ SNR $>= 5$  &1.6 & 2.2\\
SDSS spec match within 1$\arcsec$ of AllWISE & 0.00 & 0.01\\
 \hline
\end{tabular}
\end{center}
 Misidentification rates calculated by randomly shifting sources in two regions, as described in the text. The rate shown is the median number in the shifted sample that satisfy the criteria,  divided by the number in an unshifted  sample that satisfy the criteria.
\label{table:misidentificationRates}
\end{table}

\section{Redshift Distribution}

\subsection{The Large Sample of Radio Sources}


There are three sources in our initial large sample with redshifts of $z \geq 5$. 

When we examined the SDSS spectra for them, we found that in two cases (SDSS J105631.94-01145.1 and SDSS J111036.32+481752.3), the redshift depended almost entirely on a single line which was identified as Lyman-$\alpha$, but which had no corroborating evidence, such as a Lyman break, and there were other strong unidentified lines, making it resemble the spectrum of a low-redshift galaxy.
 We thus chose to exclude these two sources from further consideration, reducing the size of our initial large sample to 2519 sources, 
 which we call our ``large sample''. The sample of 108 IFRS is not affected by this.
 
 The large sample of 2519 sources is available as supplementary information in the online version of this paper, and a sample of the first few rows of this Table is shown in Table \ref{table:sample}.
 
 \begin{table}
\caption{The first 5 and the last 5 rows of our large sample. The full table is available as supplementary information in the online version of this paper.}
\begin{center}
 \begin{tabular}{||rrrrrrr||} 
 \hline
 ID & RA & Dec & $S_{20cm}$ & $S_{3.4\mu m}$ & z & $\delta$z
   \\ [0.5ex] 
 \hline\hline

1	&	221.68050	&	27.95017	&	1113.9	&	303.3	&	0.00692	&	0.00007	\\
2	&	229.18569	&	7.02204	&	5499.3	&	8635.5	&	0.03453	&	0.00001	\\
3	&	117.03940	&	30.10852	&	225.5	&	324.7	&	0.04209	&	0.00001	\\
4	&	152.00014	&	7.50458	&	6522.1	&	149.8	&	0.06813	&	0.00004	\\
5	&	9.26650	&	-1.15124	&	4067.1	&	4420.7	&	0.07365	&	0.00001	\\
2515	&	243.06980	&	47.04826	&	53.5	&	72.8	&	4.36238	&	0.00072	\\
2516	&	233.89124	&	2.90653	&	59.5	&	25.1	&	4.38719	&	0.00129	\\
2517	&	145.02005	&	5.44189	&	61.7	&	39.9	&	4.50384	&	0.00055	\\
2518	&	210.10587	&	31.81968	&	21.9	&	41.7	&	4.69133	&	0.00053	\\
2519	&	156.59841	&	25.71657	&	256.9	&	65.6	&	5.27746	&	0.00064	\\
  
 \hline
\end{tabular}
\end{center}
Column descriptions:\\
1: Unique ID from 1-2519\\
2: FIRST (Becker et al. 1995) Right Ascension (J2000) in degrees\\
3: FIRST (Becker et al. 1995) Declination (J2000) in degrees\\
4: NVSS (Condon et al. 1998) 20 cm radio flux density in mJy\\
5: AllWISE (Cutri et al. 2013) 3.4 $\mu$m infrared flux density in $\mu$Jy\\
6: SDSS DR12 (Alam et al. 2015) spectroscopic redshift\\
7: Mean SDSS DR12 (Alam et al. 2015) spectroscopic redshift error
\label{table:sample}
\end{table}
 
 
 In the remaining high-redshift galaxy, SDSS J102623.62+254259.6, the identification of the putative Lyman-$\alpha$ line was confirmed by the presence of a strong Lyman break. 
 SDSS J102623.62+254259.6 is therefore a radio-loud quasar at a redshift of 5.28, making it one of the highest-redshift radio-loud sources known.
 
Two other sources (SDSS J151656.6+183021 = 3C316, and SDSS J101115.64+010642.5 = PKS 1008+013) had large but unreliable redshifts listed by SDSS, and modest but reliable redshifts from other authors, and so their redshifts were corrected in our database.

\begin{figure*}
	\includegraphics[width=17.4cm]{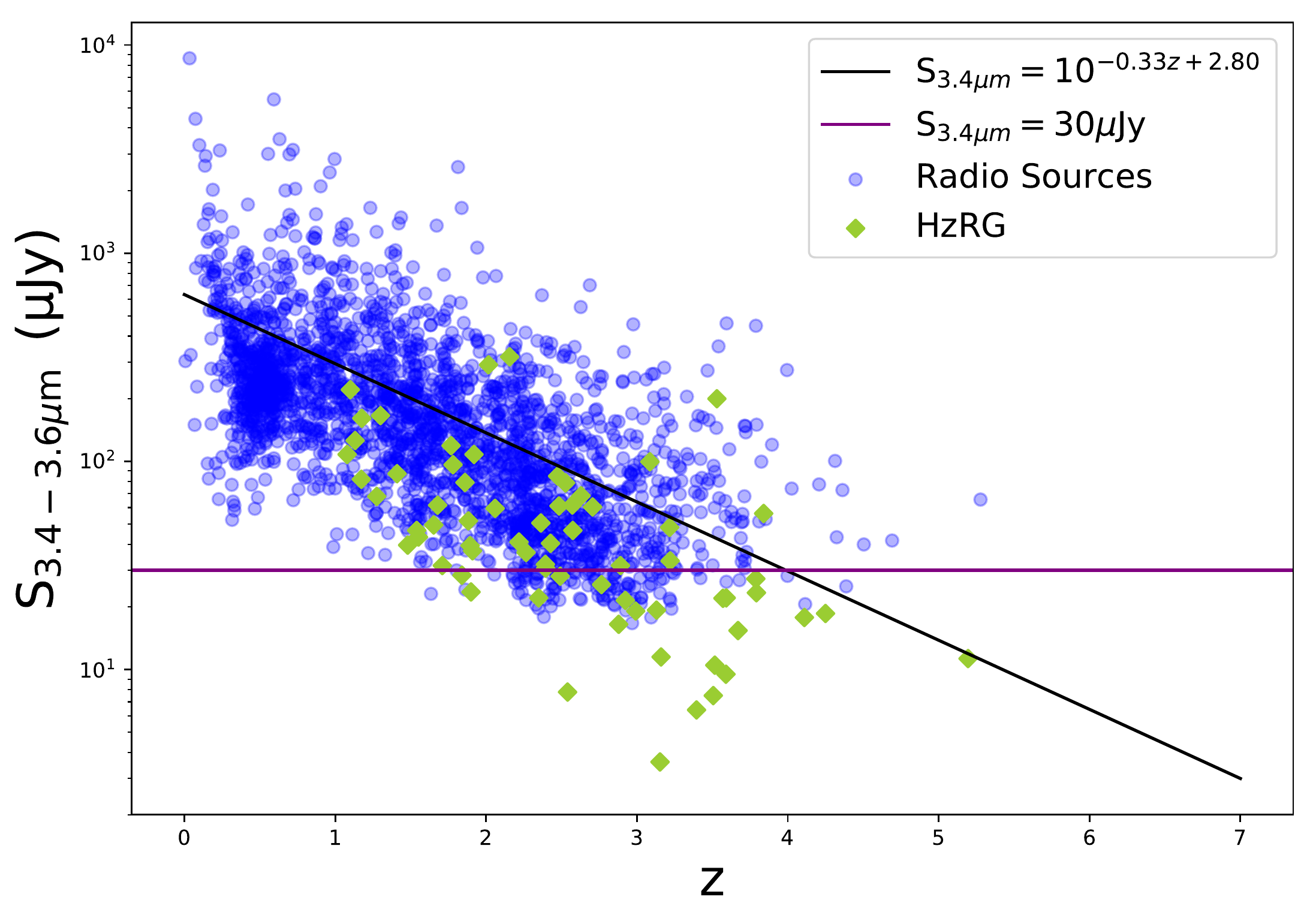}
    \caption{
The $S_{3.4-3.6{\mu}\mathrm{m}}$ 
flux density as a function of $z$, the spectroscopic redshift, for our selected radio sources, 
plotted as blue circles, and
the HzRGs from 
\citet{seymour07}, 
plotted as green diamonds. 
The sources clearly follow an IR-$z$ correlation which, as suggested by  \citet{norris11}, might yield even higher redshift 
sources by correlating radio surveys with higher sensitivity infrared surveys. Our IFRS sample consists of the 108 blue circles that lie below the horizontal line marking the 30 $\mu$Jy flux density limit.}
    \label{fig:IFRS_RadioSources}
\end{figure*}

In Figure~\ref{fig:IFRS_RadioSources} we plot the $S_{3.4{\mu}\mathrm{m}}$ flux density of the 2519 galaxies in the large sample as a function of redshift.  
This is discussed further in Section \ref{irz} below.

\begin{figure*}
	\includegraphics[width=17.4cm]{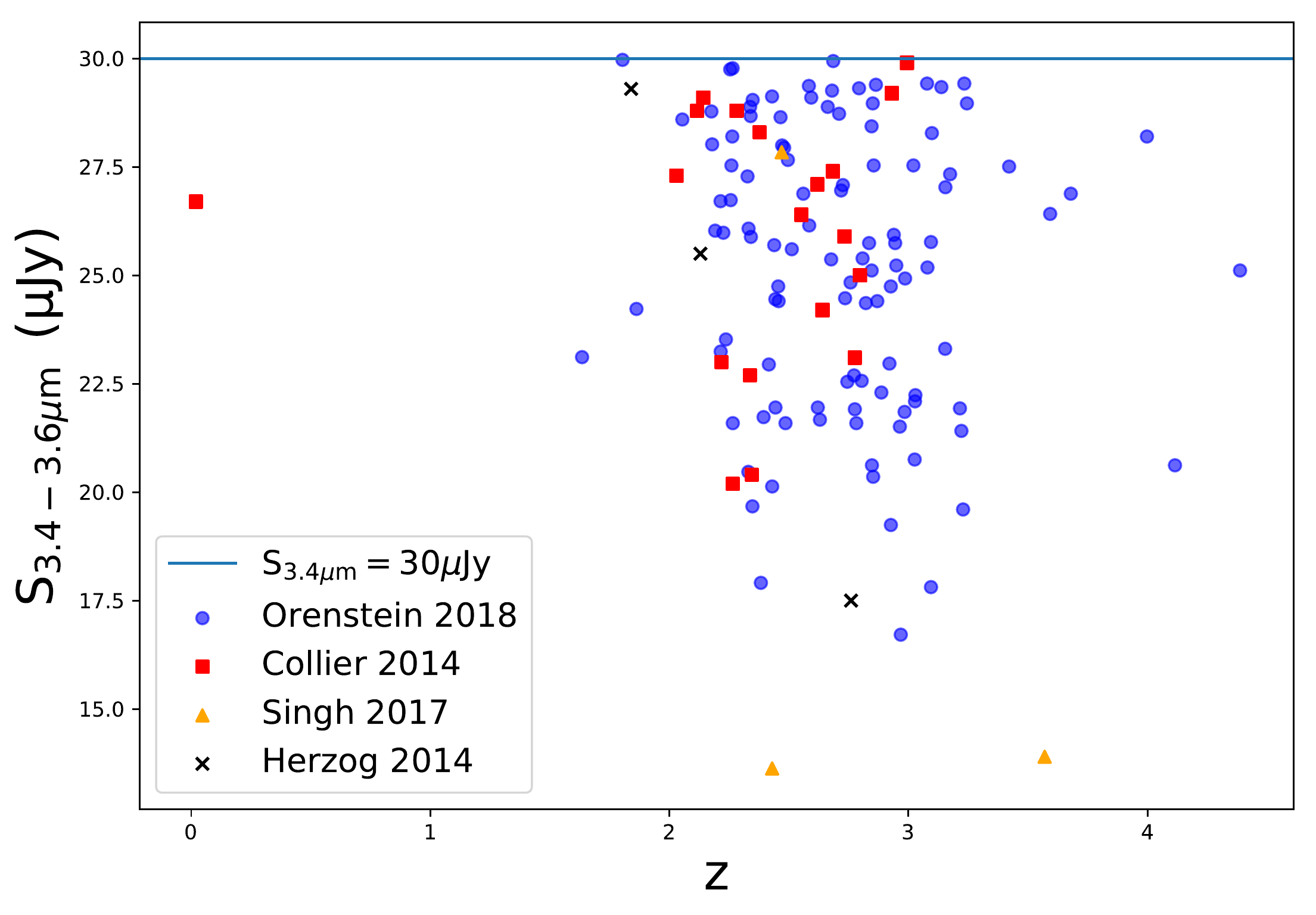}
        \caption{The 
        $S_{3.4-3.6{\mu}\mathrm{m}}$ 
        flux density as a function of $z$, the spectroscopic redshift, for our sample of 108 IFRSs, plotted as blue circles, the IFRSs from 
        \citet{collier14}, 
        plotted as red squares, the IFRSs from \citet{singh17}, plotted as orange triangles, and the IFRSs from \citet{herzog14}, plotted as black crosses. Sources from \citet{herzog15} are not shown since they are a subset of those already identified in 
        \citet{collier14}.
        }
    \label{fig:IFRSs}
\end{figure*}


\subsection{IFRSs}
After we applied the entire 14 selection criteria, our sample consisted of 108 IFRSs (106 of which are new) with spectroscopic redshifts from SDSS, listed in Table \ref{table:IFRSs}. We plot their $S_{3.4{\mu}\mathrm{m}}$ flux densities as a function of redshift in Figure~\ref{fig:IFRSs}. For comparison, we also show the IFRS samples from \cite{collier14}, \cite{herzog14}, and \cite{singh17}. 

There is a clear overlap between the  samples, and a high density of IFRSs up to the $S_{\rm 3.4\,\mu m} < 30\,\mu$Jy cutoff. Nearly all IFRSs in these samples lie between two sharply defined boundaries at about $z$ = 2.4 and $z$ = 3.4. 
However, Figure~\ref{fig:IFRS_RadioSources} shows that these boundaries are caused by the intersection of the IR-$z$ correlation and our IR flux limit.


\section{Discussion}

\subsection{The IFRS Sample}
Before the present work, only 25  IFRSs had measured spectroscopic redshifts \citep{collier14,herzog14,singh17}, with a maximum redshift of 3.570. Our  sample gives a fivefold increase in the number of IFRS with spectroscopic redshifts, adding 106 sources, and raises the maximum redshift to 4.387 (ID 86 in Table \ref{table:IFRSs}). 


\subsection{Confirming the Flux Density -- Redshift Correlation}
\label{irz}

In Figure~\ref{fig:IFRS_RadioSources} we plot the $S_{\rm 3.4\,\mu m}$ flux density of the 2519 galaxies in the large sample as a function of redshift.  
For comparison,  we also show the 69 High-$z$ Radio Galaxies (HzRGs) from ~\cite{seymour07}. It is clear from this plot that the galaxies show the inverse correlation between 3.6$\,\mu$m flux density and redshift first noted by \cite{norris11}. 

This correlation, which is similar to the well-known $k-z$ relation \citep{willott03}, was confirmed by ~\cite{singh17} for all known IFRSs and the \cite{seymour07} sample of HzRGs. With the much larger number of sources now available, we  confirmed and  refined this correlation by  fitting a straight-line to the data in log-lin space. We used the Levenberg-Marquardt algorithm \citep{more1978} to perform a nonlinear least-squares fit with SciPy \citep{jones2001}, and obtained the relation:
\begin{equation}
    S_{\rm 3.4\,{\mu}m} = 10^{-0.33(\pm 0.02)z+2.80(\pm 0.02)}
    \label{eq:fit}
\end{equation}
This is shown as the black line in Figure 
\ref{fig:IFRS_RadioSources}.

This correlation (hereafter called the IR-$z$ correlation) shows that, as suggested by \cite{norris11}, even higher redshift sources might be found by correlating radio surveys with even higher sensitivity infrared surveys.



\subsection{Redshift distribution}
\begin{figure}
	\includegraphics[width=\columnwidth]{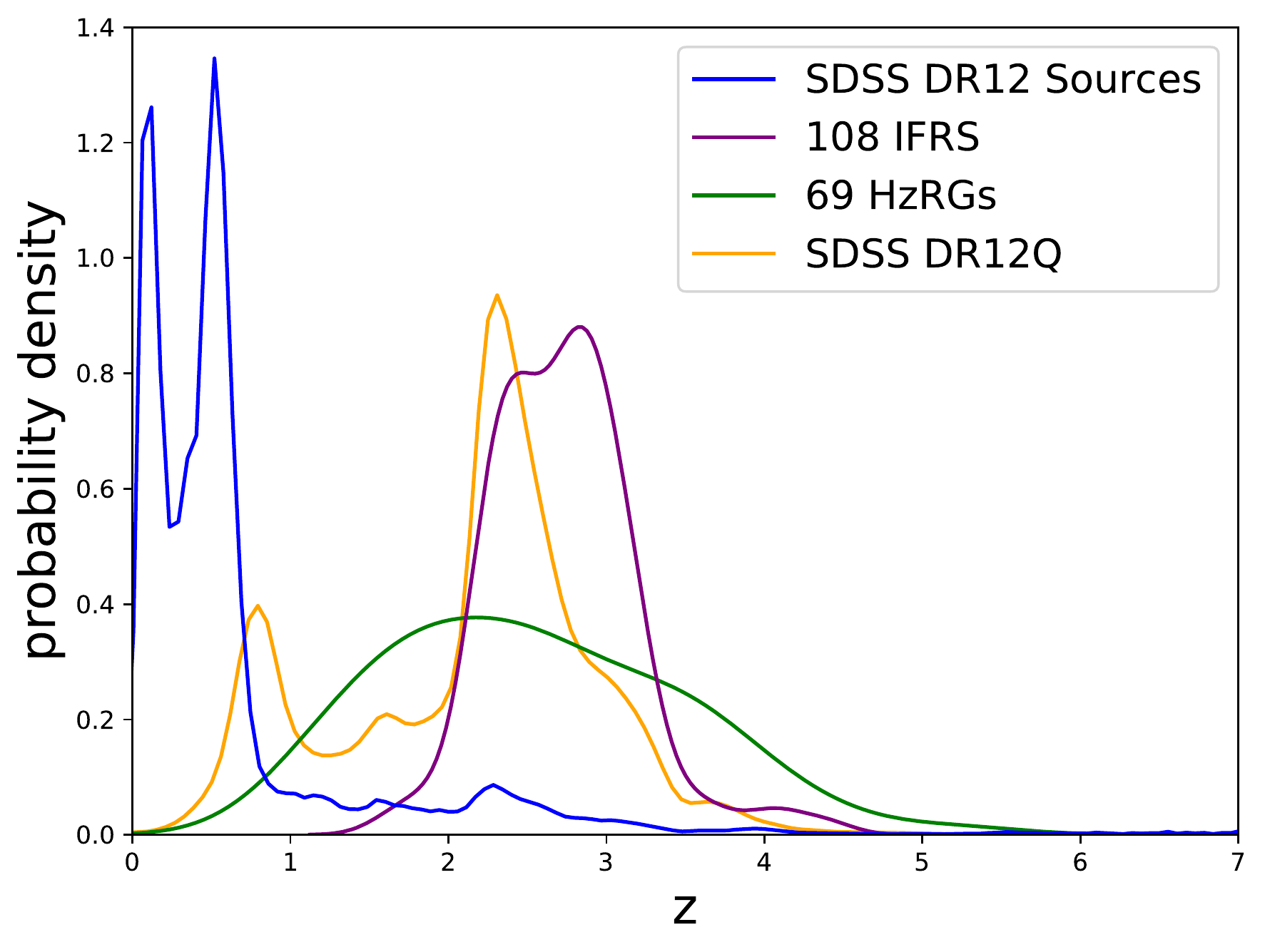}
    \label{fig:IFRSsKDE}
\end{figure}

In Figure~\ref{fig:IFRSsKDE} we plot the redshift distribution of the HzRGs from \cite{seymour07}, the SDSS Quasar Catalog twelfth data release (SDSS DR12Q) pipeline redshift estimates \cite{paris17}, our IFRSs and a random 10 per cent of SDSS DR12 sources with spectroscopic redshifts.
The HzRGs have a median redshift of $z=2.13$, and a wide redshift distribution consistent with that noted by \cite{seymour07}.
 
The IFRSs in our sample have a narrower redshift distribution, with a median redshift of \textbf{$z=2.68$.}   
As would be expected, the majority of the SDSS sample contains lower redshifts, with a median redshift of $z=0.32$ and 87.50 per cent of the redshifts less than 1, clearly eliminating the possibility that IFRSs are misidentifications. 


Figures \ref{fig:IFRS_RadioSources} and \ref{fig:IFRSs} show that the redshift distribution of our IFRS sample is defined by the intersection of the IR-$z$ correlation with our IR flux density limit.

\subsection{Using IFRSs to find high-z radio sources}
Obtaining a census of high-redshift radio sources is important for several reasons. 

First, conventional hierarchical models of the formation of super-massive black holes (SMBHs) are unable to produce such high-mass black holes early in the lifetime of the Universe, driving the development of novel models for supermassive black hole formation \citep[e.g.][]{volonteri05}. HzRGs represent a subset of high-redshift super-massive black holes that are relatively easy to detect in large radio surveys, although measuring their redshifts is  challenging. Finding a significant number of HzRGs at $z>6$ would exacerbate the problem of the formation of SMBHs at high redshift.


Second, high redshift radio sources are important cosmological probes, as they provide background sources against which HI absorption may be seen at high redshifts \citep{carilli02, ciardi15}

If radio galaxies continue to follow the IR-$z$ correlation shown in Figure \ref{fig:IFRS_RadioSources} to high redshift, then in principle HzRGs may be found by searching for radio sources with low IR flux densities. 
Such sources may already exist in current radio catalogues, and a great many more will be available from next-generation radio surveys \citep{norris17} such as
Evolutionary Map of the Universe \citep[EMU;][]{norris11a},
MeerKAT International GHz Tiered Extragalactic Exploration Survey \citep[MIGHTEE;][]{jarvis18},
the VLA Sky Survey\citep[VLASS;][]{murphy15},
and the	LOFAR	Two-metre	Sky	Survey \citep[LoTSS;][]{shimwell17}.

However, there are two potential challenges to be overcome to achieve this.

First, the sample shown in Figure 3 is limited by the sensitivity of the WISE survey. To detect sources at higher $z$, we need more sensitive infrared data. The SERVS survey \citep{mauduit12} reaches an r.m.s. sensitivity at 3.6$\,\mu$m of about 0.2 $\mu$Jy, so that 5$\sigma$ SERVS detections should extend to about $z=7$. Because of strong radio source evolution \citep[e.g.][]{norris13}, such sources may still be well above the radio detection limit.

However, the SERVS survey only covers a few tens of square degrees, limiting the number of high-$z$ IFRSs that can be detected \citep{maini16}. No other infrared survey in the near future will provide the required sensitivity over a large area of sky, although it is possible that $K$-band surveys such as the Vista Hemisphere Survey \citep[VHS;][]{mcmahon13}, 
UKIRT Infrared Deep Sky Survey \citep[UKIDSS;][]{lawrence07} and VISTA Kilo-degree Infrared Galaxy Survey \citep[VIKING;][]{edge13}  may be used for the same purpose together with corresponding radio surveys.

Second, we need to measure redshifts for these sources. Future large spectroscopic surveys such as
4MOST \citep{quirrenbach15},
WEAVE-LOFAR \citep{smith16},
DESI \citep{levi13},
and PFS \citep{takada14}
may be able to provide this data, but the faintest sources will still be difficult to access at optical wavelengths. 
Potential alternatives include photometric redshift techniques, although these will be limited by the faintness of the optical/IR emission, or using blind scans for CO emission using ALMA \citep[e.g.][]{weiss13}.


\section{Conclusion}
From cross-matching FIRST, AllWISE and SDSS DR12, we have compiled a large sample of 108 IFRSs with spectroscopic redshift measurements, greatly extending the previous sample of 25.  This new sample includes the highest identified IFRS spectroscopic redshift with $z=4.387$. 
This sample will be valuable in future multi-wavelength studies, with high-angular-resolution radio imaging, to study the properties of IFRSs and determine what they can tell us about the evolution of AGN.

We have also shown that IFRSs, as well as a large sample of 2519 high-redshift radio sources with spectroscopic redshifts, follow a correlation of 3.4$\,\mu$m flux density as a function of redshift, with the form  $    S_{\rm 3.4\,{\mu}m} = 10^{-0.33(\pm 0.02)z+2.80(\pm 0.02)}$. By extending this correlation to even fainter mid-infrared flux densities, this correlation appears to be a powerful tool for finding high-redshift radio sources, rivalling the use of steep-spectral index.


Future detections of high-redshift radio sources will be important for probing the high-redshift Universe, and to understand SMBH formation and AGN evolution. Future large radio surveys are likely to yield many such sources, provided that we have matching deep infrared photometry and a means of measuring redshifts.

\section{Acknowledgements}
We thank Dr. Rob Sharp for invaluable advice on SDSS spectroscopy, and Prof. Tom Jarrett for help with interpreting WISE flux densities.

We gratefully acknowledge the people and institutes that contributed to the NVSS and FIRST surveys.

This research has made use of Version 2.0 of the Unified Radio Catalog,~\citep[URC;][]{kimball08,kimball14}. The URC v2.0 website is \href{http://www.aoc.nrao.edu/~akimball/radiocat_2.0.shtml}{http://www.aoc.nrao.edu/{\textasciitilde}akimball/radiocat\_2.0.shtml}.

This research has made extensive use of the NASA/IPAC Extragalactic Database (NED) which is operated by the Jet Propulsion Laboratory, California Institute of Technology, under contract with the National Aeronautics and Space Administration.

This research has made use of the VizieR catalog access tool, CDS, Strasbourg, France. Topcat \citep{2005ASPC..347...29T}, NASA's Astrophysics Data System bibliographic services, and Astropy, a community-developed core Python package for Astronomy \citep{2018arXiv180102634T}, were also used in this study.

This research made use of APLpy, an open-source plotting package for Python \citep{2012ascl.soft08017R}.

This research made use of the Jupyter Notebook environment \citep{kluyver2016} and the Seaborn data visualisation library \citep{waskom2014}.

This research made use of the following software in the SciPy \citep{jones2001} ecosystem: Numpy \citep{oliphant2015}, iPython \citep{perez2007}, Matplotlib \citep{hunter2007} and Pandas \citep{mckinney2010}.

This research made use of Montage. It is funded by the National Science Foundation under Grant Number ACI-1440620, and was previously funded by the National Aeronautics and Space Administration's Earth Science Technology Office, Computation Technologies Project, under Cooperative Agreement Number NCC5-626 between NASA and the California Institute of Technology.

Funding for the Sloan Digital Sky Survey IV has been provided by the Alfred P. Sloan Foundation, the U.S. Department of Energy Office of Science, and the Participating Institutions. SDSS-IV acknowledges
support and resources from the Center for High-Performance Computing at
the University of Utah. The SDSS website is \href{https://www.sdss.org}{https://www.sdss.org}.

SDSS-IV is managed by the Astrophysical Research Consortium for the 
Participating Institutions of the SDSS Collaboration including the 
Brazilian Participation Group, the Carnegie Institution for Science, 
Carnegie Mellon University, the Chilean Participation Group, the French Participation Group, Harvard-Smithsonian Center for Astrophysics, 
Instituto de Astrof\'isica de Canarias, The Johns Hopkins University, 
Kavli Institute for the Physics and Mathematics of the Universe (IPMU) / 
University of Tokyo, Lawrence Berkeley National Laboratory, 
Leibniz Institut f\"ur Astrophysik Potsdam (AIP),  
Max-Planck-Institut f\"ur Astronomie (MPIA Heidelberg), 
Max-Planck-Institut f\"ur Astrophysik (MPA Garching), 
Max-Planck-Institut f\"ur Extraterrestrische Physik (MPE), 
National Astronomical Observatories of China, New Mexico State University, 
New York University, University of Notre Dame, 
Observat\'ario Nacional / MCTI, The Ohio State University, 
Pennsylvania State University, Shanghai Astronomical Observatory, 
United Kingdom Participation Group,
Universidad Nacional Aut\'onoma de M\'exico, University of Arizona, 
University of Colorado Boulder, University of Oxford, University of Portsmouth, 
University of Utah, University of Virginia, University of Washington, University of Wisconsin, 
Vanderbilt University, and Yale University.

This publication makes use of data products from the Wide-field Infrared Survey Explorer, which is a joint project of the University of California, Los Angeles, and the Jet Propulsion Laboratory/California Institute of Technology, and NEOWISE, which is a project of the Jet Propulsion Laboratory/California Institute of Technology. WISE and NEOWISE are funded by the National Aeronautics and Space Administration.





\medskip
\bibliography{main} 
\bibliographystyle{mnras}



\clearpage
\onecolumn


\begin{longtable}{cccccccc}
\caption{Our sample of 108 IFRSs with spectroscopic redshift measurements}\\
\label{table:IFRSs}
ID & SDSS ID & FIRST RA & FIRST Dec & $S_\mathrm{20\,cm}$ & $S_{3.4\,\mathrm{{\mu}m}}$ & $z$ & $\pm z$  \\ 
& & (J2000) & (J2000) &
 (mJy) & ($\mathrm{{\mu}Jy}$) & & ($10^{-3}$) \\
\hline
\hline
1   &  J004842.69+000543.7 &   12.17789 &   0.09539 &                        30.69 &                                 29.94 &  2.69 &                 0.27 \\
2   &  J012753.69+002516.4 &   21.97376 &   0.42131 &                        91.46 &                                 24.75 &  2.46 &                 0.10 \\
3   &  J014934.63-024305.3 &   27.39430 &  -2.71818 &                        20.26 &                                 26.03 &  2.19 &                 0.70 \\
4   &  J020553.54-001338.7 &   31.47305 &  -0.22738 &                        13.42 &                                 22.69 &  2.77 &                 0.78 \\
5   &  J025139.59-083432.5 &   42.91476 &  -8.57575 &                        69.12 &                                 24.75 &  2.93 &                 0.77 \\
6   &  J025808.78-020912.7 &   44.53675 &  -2.15349 &                        17.99 &                                 28.67 &  2.34 &                 0.41 \\
7   &  J073459.45+420425.4 &  113.74677 &  42.07347 &                        20.66 &                                 26.74 &  2.26 &                 0.55 \\
8   &  J074013.97+463853.9 &  115.05820 &  46.64835 &                        27.67 &                                 28.59 &  2.05 &                 1.20 \\
9   &  J074105.87+294909.7 &  115.27448 &  29.81936 &                        33.50 &                                 25.74 &  2.84 &                 0.32 \\
10  &  J074950.69+191152.5 &  117.46117 &  19.19794 &                        19.91 &                                 26.88 &  2.56 &                 0.14 \\
11  &  J080541.11+512707.4 &  121.42130 &  51.45205 &                        27.72 &                                 26.15 &  2.59 &                 0.54 \\
12  &  J081553.19+063858.4 &  123.97165 &   6.64958 &                        25.03 &                                 27.51 &  3.42 &                 0.46 \\
13  &  J081948.76+452434.1 &  124.95320 &  45.40952 &                        34.08 &                                 25.39 &  2.81 &                 0.26 \\
14  &  J082617.09+362115.6 &  126.57134 &  36.35435 &                        18.64 &                                 28.78 &  2.18 &                 0.45 \\
15  &  J083221.85+313518.4 &  128.09043 &  31.58768 &                        54.53 &                                 22.57 &  2.80 &                 0.80 \\
16  &  J083935.95+011214.5 &  129.89975 &   1.20408 &                        17.50 &                                 28.44 &  2.85 &                 0.52 \\
17  &  J083955.38+025145.4 &  129.98063 &   2.86267 &                        78.60 &                                 26.88 &  3.68 &                 0.23 \\
18  &  J084423.07+523920.3 &  131.09626 &  52.65560 &                        12.54 &                                 22.24 &  3.03 &                 0.39 \\
19  &  J085157.78+442107.9 &  132.99075 &  44.35220 &                        12.87 &                                 24.93 &  2.99 &                 0.30 \\
20  &  J090259.94+272028.2 &  135.74976 &  27.34119 &                        83.35 &                                 29.37 &  2.58 &                 0.30 \\
21  &  J093450.93+460329.0 &  143.71228 &  46.05806 &                        13.46 &                                 23.52 &  2.24 &                 0.50 \\
22  &  J093536.18+291710.8 &  143.90071 &  29.28636 &                        50.01 &                                 28.00 &  2.47 &                 0.62 \\
23  &  J094523.07+365555.5 &  146.34615 &  36.93206 &                        36.54 &                                 29.13 &  2.43 &                 0.42 \\
24  &  J094705.51+302008.5 &  146.77299 &  30.33567 &                        47.71 &                                 25.77 &  3.09 &                 0.48 \\
25  &  J095043.72+594631.2 &  147.68208 &  59.77535 &                        11.71 &                                 19.60 &  3.23 &                 0.28 \\
26  &  J100653.88+595519.6 &  151.72447 &  59.92216 &                        14.80 &                                 20.36 &  2.85 &                 0.45 \\
27  &  J100655.80+050324.8 &  151.73256 &   5.05686 &                        29.57 &                                 29.42 &  3.08 &                 0.71 \\
28  &  J101032.23+080805.2 &  152.63407 &   8.13474 &                        20.49 &                                 27.28 &  2.33 &                 0.68 \\
29  &  J102503.59+390350.1 &  156.26502 &  39.06389 &                        26.97 &                                 24.45 &  2.44 &                 0.53 \\
30  &  J102823.47+500913.9 &  157.09777 &  50.15408 &                        14.30 &                                 26.96 &  2.72 &                 0.97 \\
31  &  J102846.94+412656.7 &  157.19568 &  41.44908 &                        71.99 &                                 24.36 &  2.82 &                 0.54 \\
32  &  J111141.06+562503.5 &  167.92120 &  56.41775 &                        65.04 &                                 25.60 &  2.51 &                 0.45 \\
33  &  J111636.11+583231.0 &  169.15048 &  58.54193 &                        19.92 &                                 27.54 &  2.85 &                 0.34 \\
34  &  J111658.03+520333.5 &  169.24183 &  52.05913 &                        29.56 &                                 28.89 &  2.34 &                 0.30 \\
35  &  J112341.85+091328.4 &  170.92440 &   9.22458 &                        20.08 &                                 29.75 &  2.25 &                 0.37 \\
36  &  J112344.71+342546.7 &  170.93641 &  34.42975 &                        14.75 &                                 27.03 &  3.15 &                 0.25 \\
37  &  J112356.35+462901.2 &  170.98462 &  46.48382 &                        31.36 &                                 26.71 &  2.21 &                 0.45 \\
38  &  J112549.64+482759.6 &  171.45682 &  48.46654 &                        24.48 &                                 28.20 &  4.00 &                 0.40 \\
39  &  J113116.45+514634.2 &  172.81872 &  51.77623 &                        93.67 &                                 25.74 &  2.94 &                 0.37 \\
40  &  J113605.23+222218.2 &  174.02162 &  22.37173 &                        32.83 &                                 28.28 &  3.10 &                 0.18 \\
41  &  J113610.45+314924.9 &  174.04336 &  31.82377 &                        15.26 &                                 29.31 &  2.79 &                 0.77 \\
42  &  J113902.81+231016.7 &  174.76174 &  23.17148 &                        44.45 &                                 28.65 &  2.47 &                 0.26 \\
43  &  J113904.76+245712.1 &  174.76979 &  24.95311 &                        30.08 &                                 28.20 &  2.26 &                 0.36 \\
44  &  J115428.30+141004.2 &  178.61794 &  14.16786 &                        14.70 &                                 29.26 &  2.68 &                 0.29 \\
45  &  J115650.86+353103.5 &  179.21206 &  35.51766 &                       197.38 &                                 29.34 &  3.14 &                 0.59 \\
46  &  J121129.17+243958.9 &  182.87153 &  24.66634 &                        25.00 &                                 27.54 &  3.02 &                 0.63 \\
47  &  J121230.17+251321.2 &  183.12567 &  25.22255 &                       216.33 &                                 24.41 &  2.87 &                 0.81 \\
48  &  J121840.03+244955.0 &  184.66686 &  24.83194 &                       239.19 &                                 28.97 &  2.85 &                 0.18 \\
49  &  J122046.01+494508.4 &  185.19181 &  49.75238 &                        14.94 &                                 24.84 &  2.76 &                 2.00 \\
50  &  J124541.71+255918.0 &  191.42293 &  25.98868 &                        15.02 &                                 29.78 &  2.26 &                 0.24 \\
51  &  J125134.74+581257.6 &  192.89482 &  58.21599 &                        13.79 &                                 19.24 &  2.93 &                 1.07 \\
52  &  J125230.57+245813.3 &  193.12733 &  24.97042 &                        14.27 &                                 27.94 &  2.48 &                 0.66 \\
53  &  J125300.15+524803.3 &  193.25067 &  52.80096 &                        58.02 &                                 20.62 &  4.12 &                 0.53 \\
54  &  J125656.83+385540.2 &  194.23691 &  38.92792 &                        18.08 &                                 21.95 &  2.44 &                 0.70 \\
55  &  J125832.23+454305.1 &  194.63454 &  45.71699 &                        13.56 &                                 22.95 &  2.42 &                 0.90 \\
56  &  J130417.79+564729.9 &  196.07378 &  56.79157 &                        12.32 &                                 20.13 &  2.43 &                 0.46 \\
57  &  J130744.40+212615.5 &  196.93514 &  21.43769 &                        98.70 &                                 24.41 &  2.46 &                 0.31 \\
58  &  J132622.76+582750.1 &  201.59474 &  58.46398 &                        44.54 &                                 21.85 &  2.98 &                 0.84 \\
59  &  J133202.45+211317.8 &  203.01015 &  21.22149 &                        14.56 &                                 27.33 &  3.17 &                 0.31 \\
60  &  J133221.80-022449.7 &  203.09031 &  -2.41391 &                        29.54 &                                 28.97 &  3.24 &                 0.35 \\
61  &  J133714.92+352958.5 &  204.31208 &  35.49965 &                        33.72 &                                 28.73 &  2.71 &                 0.43 \\
62  &  J134329.90+320800.2 &  205.87462 &  32.13342 &                        19.54 &                                 23.31 &  3.15 &                 0.30 \\
63  &  J134637.43+290042.4 &  206.65597 &  29.01181 &                        36.48 &                                 27.08 &  2.73 &                 0.15 \\
64  &  J140828.37+060902.9 &  212.11765 &   6.15066 &                        19.43 &                                 25.89 &  2.34 &                 0.33 \\
65  &  J141204.71+282237.7 &  213.01962 &  28.37714 &                        35.16 &                                 27.66 &  2.50 &                 0.17 \\
66  &  J141309.25+240700.6 &  213.28852 &  24.11682 &                        20.56 &                                 29.05 &  2.35 &                 0.49 \\
67  &  J141529.73-021448.2 &  213.87350 &  -2.24708 &                        16.07 &                                 23.24 &  2.21 &                 0.29 \\
68  &  J141637.23+143044.2 &  214.15515 &  14.51232 &                       134.11 &                                 21.59 &  2.78 &                 1.44 \\
69  &  J143130.84+233422.2 &  217.87853 &  23.57281 &                        23.62 &                                 26.42 &  3.59 &                 0.37 \\
70  &  J143243.17+232009.4 &  218.17992 &  23.33607 &                        14.66 &                                 22.30 &  2.89 &                 0.67 \\
71  &  J143301.51+010132.9 &  218.25632 &   1.02586 &                        40.53 &                                 29.42 &  3.23 &                 0.32 \\
72  &  J143909.90+485950.6 &  219.79122 &  48.99744 &                        10.63 &                                 20.62 &  2.85 &                 0.75 \\
73  &  J144514.18+414102.4 &  221.30921 &  41.68406 &                        34.53 &                                 22.09 &  3.03 &                 0.33 \\
74  &  J144837.56-025036.5 &  222.15661 &  -2.84354 &                        14.73 &                                 23.11 &  1.63 &                 0.58 \\
75  &  J145207.22+201100.6 &  223.02983 &  20.18425 &                        23.25 &                                 25.70 &  2.44 &                 0.73 \\
76  &  J145627.56+435500.0 &  224.11490 &  43.91671 &                        27.21 &                                 25.23 &  2.95 &                 0.45 \\
77  &  J145902.70+495535.6 &  224.76134 &  49.92657 &                        24.34 &                                 29.10 &  2.59 &                 1.16 \\
78  &  J150048.62+452805.7 &  225.20276 &  45.46843 &                        17.42 &                                 24.47 &  2.74 &                 0.33 \\
79  &  J150239.02+172145.2 &  225.66251 &  17.36262 &                        20.85 &                                 22.97 &  2.92 &                 0.69 \\
80  &  J150419.14+214112.0 &  226.07931 &  21.68623 &                        45.98 &                                 27.54 &  2.26 &                 0.37 \\
81  &  J151609.84+222507.7 &  229.04101 &  22.41884 &                        20.08 &                                 21.91 &  2.78 &                 0.82 \\
82  &  J151851.18+334821.9 &  229.71331 &  33.80606 &                        28.48 &                                 20.47 &  2.33 &                 0.23 \\
83  &  J152512.18+250653.0 &  231.30077 &  25.11470 &                        23.15 &                                 20.75 &  3.03 &                 0.43 \\
84  &  J152615.08+245425.6 &  231.56291 &  24.90710 &                        40.97 &                                 25.37 &  2.68 &                 0.22 \\
85  &  J152926.77+404004.7 &  232.36139 &  40.66788 &                        55.21 &                                 17.81 &  3.09 &                 0.57 \\
86  &  J153533.88+025423.3 &  233.89124 &   2.90653 &                        59.49 &                                 25.11 &  4.39 &                 1.29 \\
87  &  J153607.55+162800.9 &  234.03152 &  16.46713 &                        53.59 &                                 29.40 &  2.86 &                 1.21 \\
88  &  J153957.12+095503.5 &  234.98806 &   9.91763 &                        20.99 &                                 25.11 &  2.85 &                 0.85 \\
89  &  J154105.41+292231.0 &  235.27260 &  29.37533 &                        19.27 &                                 25.94 &  2.94 &                 0.82 \\
90  &  J154314.71+325138.1 &  235.81155 &  32.86052 &                        50.28 &                                 21.59 &  2.27 &                 0.32 \\
91  &  J154409.51+082425.5 &  236.03942 &   8.40675 &                        18.41 &                                 26.08 &  2.33 &                 0.24 \\
92  &  J154516.07+504253.3 &  236.31687 &  50.71481 &                        32.67 &                                 22.55 &  2.74 &                 0.27 \\
93  &  J155252.25+425929.6 &  238.21739 &  42.99161 &                        16.35 &                                 16.71 &  2.97 &                 0.96 \\
94  &  J155539.00+233014.8 &  238.91215 &  23.50444 &                        22.86 &                                 24.23 &  1.86 &                 0.74 \\
95  &  J160601.45+293150.9 &  241.50609 &  29.53085 &                        16.88 &                                 28.02 &  2.18 &                 0.20 \\
96  &  J160820.89+183059.5 &  242.08707 &  18.51661 &                        22.39 &                                 21.41 &  3.22 &                 0.29 \\
97  &  J160856.08+091038.5 &  242.23371 &   9.17729 &                        23.26 &                                 21.93 &  3.22 &                 0.91 \\
98  &  J161330.32+404423.3 &  243.37627 &  40.73996 &                        24.02 &                                 19.67 &  2.35 &                 0.50 \\
99  &  J161800.87+542645.0 &  244.50426 &  54.44600 &                        11.85 &                                 21.51 &  2.96 &                 0.67 \\
100 &  J162347.43+544302.7 &  245.94758 &  54.71748 &                        22.76 &                                 25.98 &  2.23 &                 0.48 \\
101 &  J163241.76+514955.3 &  248.17413 &  51.83202 &                        15.26 &                                 17.91 &  2.38 &                 0.49 \\
102 &  J163355.47+254126.4 &  248.48117 &  25.69068 &                        29.75 &                                 21.95 &  2.62 &                 1.37 \\
103 &  J163708.44+273143.0 &  249.28401 &  27.52820 &                        13.55 &                                 21.59 &  2.49 &                 0.49 \\
104 &  J163946.92+403933.4 &  249.94617 &  40.65971 &                        19.75 &                                 21.67 &  2.63 &                 0.26 \\
105 &  J164212.29+191848.0 &  250.55116 &  19.31342 &                        12.69 &                                 21.73 &  2.39 &                 0.87 \\
106 &  J164524.88+224508.0 &  251.35370 &  22.75222 &                        17.48 &                                 25.18 &  3.08 &                 0.74 \\
107 &  J164612.06+303202.2 &  251.54944 &  30.53358 &                        37.63 &                                 29.97 &  1.80 &                 0.71 \\
108 &  J211742.96+061126.4 &  319.42908 &   6.19078 &                        18.16 &                                 28.89 &  2.66 &                 0.24 \\
\hline
\end{longtable}
\bsp	
\label{lastpage}
\end{document}